\documentclass[twocolumn,showpacs,preprintnumbers,amsmath,amssymb,pre]{revtex4}

\usepackage{epsfig,amsmath,amssymb,graphics,color,calc}

\begin{document}
\title{Direct-space investigation of the ultraslow ballistic dynamics of a soft glass}
\author{Sylvain Mazoyer, Luca Cipelletti and Laurence Ramos$^*$}
\affiliation{Laboratoire des Collo\"{\i}des, Verres et
Nanomat\'{e}riaux (UMR CNRS-UM2 5587), CC26, Universit\'{e}
Montpellier 2, 34095 Montpellier Cedex 5, France}

\email{ramos@lcvn.univ-montp2.fr}
\date{\today}

\begin{abstract}

We use light microscopy to investigate the aging dynamics of a glass
made of closely packed soft spheres, following a rapid transition
from a fluid to a solid-like state. By measuring time-resolved,
coarse-grained displacements fields, we identify two classes of
dynamical events, corresponding to reversible and irreversible
rearrangements, respectively. The reversible events are
due to the small, experimentally unavoidable fluctuations of the
temperature imposed to the sample, leading to transient thermal
expansions and contractions that cause shear deformations. The
irreversible events are plastic rearrangements, induced by the
repeated shear cycles. We show that the displacement due to the
irreversible rearrangements grows linearly with time, both on
average and at a local level. The velocity associated with this
ballistic motion decreases exponentially with sample age, accounting
for the observed slowing down of the dynamics. The displacement
field due to the irreversible rearrangements has a vortex-like
structure and is spatially correlated over surprisingly long
distances.
\end{abstract}


\maketitle


\section{INTRODUCTION}

A large variety of soft disordered materials, ranging from colloidal
pastes and gels, hard-spheres suspensions, dense surfactant,
copolymer or star polymer systems, exhibit very slow dynamics, often
associated with physical aging, the continuous growth of relaxation
times after the sample has been prepared~\cite{CipellettiJPCM2005}.
In this respect, these systems display intriguing analogies with
molecular or polymer glasses~\cite{Donth2001} and are often termed
``Soft Glassy Materials''~\cite{SollichPRL2007}. Analogies are also
found with granular
materials~\cite{Pouliquen,Dauchot,ReisPRL2007,DurianNatPhys2007}
close to the jamming~\cite{LiuNature1998} transition: these
similarities have spurred a large research activity, aiming at
rationalizing the glass and jamming transitions within one single
conceptual framework~\cite{LiuNature1998}. Note however that the
driving force for the slow dynamics is different in molecular and
colloidal systems, where thermal energy dominates, and in athermal
granular media, where some form of external drive is required to
induce structural rearrangements.

Among soft glasses, a class of systems have emerged whose dynamics
exhibit peculiar features not seen neither in molecular glass
formers nor in granular media~\cite{LucaPRL2000,KnaebelEPJL2000,
RamosPRL2001,LucaFaraday2003,BandyopadhyayPRL2004,RobertEPL2006,
BandyopadhyayReview2006,ChungPRL2006,FalusPRL2006,GuoPRE2007,CaronnaPRL2008}.
They range from tenuous, diluted gels made of attractive
particles, to repulsive colloidal glasses or concentrated
surfactant phases. In these systems, correlation functions
measured by dynamic light scattering or X-photon correlation
spectroscopy decay steeper than exponentially (``compressed
exponential'') and their decay time scales as the inverse
scattering vector, $q$, rather than as $q^{-2}$, as expected for a
diffusive process. This suggests that the displacement of the
scatterers grows linearly with time, e.g. that the slow dynamics
is, on average, ballistic. Recent time-resolved light scattering
experiments~\cite{DuriEPL2006} on a tenuous colloidal gel, the
first system for which these unusual dynamics were
reported~\cite{LucaPRL2000}, have shown that particle
displacements are not continuous but rather result from
intermittent rearrangements that affect a very large region of the
sample~\cite{ThesisAgnes}.

In spite of the large number of scattering experiments reporting
ultraslow ballistic relaxations, the ballistic nature of the
dynamics and its discontinuous behavior have never been
demonstrated by directly imaging the motion of particles. Direct
space information~\cite{PrasadJPCM2007} on the dynamics would be
very valuable in order to better understand the physical origin of
such unusual dynamics. It is generally proposed that ballistic
dynamics stem from the relaxation of internal
stress~\cite{BouchaudEPJE2001,LucaFaraday2003}. While a plausible
microscopic picture of the mechanisms through which stress may be
accumulated and released has been proposed (but not directly
demonstrated) for tenuous colloidal gels, a convincing explanation
for concentrated systems is still lacking. In particular, it is
unclear whether thermal energy alone could be responsible for
significant restructuring in very dense colloidal or surfactant
systems, where the elastic modulus can be a factor of $10^3 -
10^4$ larger than that of the tenuous gels.

To address some of these issues, we have recently used light
microscopy to investigate the slow dynamics of a soft glass made
of a compact arrangement of elastic spheres
(``onions'')~\cite{MazoyerPRL2006}, the same system for which
ballistic dynamics were reported in previous light scattering
experiments~\cite{RamosPRL2001}. In Ref.~\cite{MazoyerPRL2006} we
used a coarse-grained method to measure two-time displacement
fields in direct space. The main conclusion of this work was that
the slow dynamics is due to shear deformations resulting from the
experimentally unavoidable small fluctuations of the temperature
imposed to the sample. Indeed, as temperature fluctuates the
sample undergoes a series of contractions and elongations, because
of thermal expansion. While macroscopically these deformations are
fully reversible, microscopically they eventually induce
irreversible rearrangements. These experiments suggested that the
slow dynamics of dense colloidal systems may share more
similarities than expected with sheared athermal systems. In
Ref.~\cite{MazoyerPRL2006} we mainly focussed on average
quantities, such as the mean squared displacement and its age
dependence. Here, we extend this study and provide a novel and
more direct method to quantify the irreversible rearrangements,
showing unambiguously that they lead to ballistic motion, both on
average and at a local level. Furthermore, we investigate in
detail the spatial structure of the irreversible rearrangements.
We find that the displacement field is correlated over distances
comparable to the system size, presenting interesting analogies
with recent simulations~\cite{BritoEPL2006,BritoJSTAT2007} of
jammed spheres and experiments on very concentrated colloidal
suspensions~\cite{BallestaNatPhys2008}.

The paper is organized as follows. Section~\ref{Exp} describes the
experimental system, the measurement procedure and the image
analysis method that we have developed. In Section~\ref{Results}, we
first present the time-resolved (but spatially averaged) dynamics
and then discuss the individual trajectories and the spatial
structure of the rearrangements. In the last section, we discuss
briefly our main findings and conclude.

\section{Materials and Methods}
\label{Exp}

\subsection{Sample preparation and optical microscopy}

The structure and the phase diagram of our system have been
discussed in detail in Ref.~\cite{equilibriumonions}: here we simply
recall their main features. The sample is a surfactant lamellar
phase constituted of a regular stacking of bilayers. The bilayers
are composed of a mixture of cetylpyridinium chloride (CpCl) and
octanol (Oct) (weight ratio $\rm{CpCl/Oct}=0.95$) and diluted in
brine ($\rm{[NaCl]}= 0.2 \rm{M}$) at a weight fraction $\phi=16 \%$.
An amphiphilic copolymer, Symperonics F68 by Serva
($\rm{(EO)_{76}}-\rm{(PO)_{29}}-\rm{(EO)_{76}}$, where EO is
ethylene oxide and PO is propylene oxide) is incorporated into the
lamellar phase. At room temperature the PO block is hydrophobic and
adsorbs onto the bilayers, while the hydrophilic EO blocks remain
swollen in the aqueous solvent and decorate the surfactant bilayers.
Upon copolymer addition, we have shown both experimentally and
theoretically that a transition from a flat lamellar phase to an
onion phase occurs \cite{equilibriumonions}. The latter consists of
a dense packing of multilamellar vesicles, or onions. Due to their
polydispersity and softness, the volume fraction of the onions is
one. In the experiments reported here, the copolymer-to-bilayer
weight ratio is $0.8$  and the radius of the largest onions is about
$6 \, \mu \rm{m}$ \cite{equilibriumonions}.

The rheological properties of our samples can be tuned by changing
the temperature, $T$. Both the EO and PO blocks of the F68 copolymer
are hydrophilic at low temperature ($T \lesssim 8^{\circ} \rm{C}$): the
 copolymer is fully soluble in water and the sample is fluid.
Because the PO block becomes increasingly hydrophobic when
increasing $T$,  the sample becomes gel-like at room temperature,
where the PO block adsorbs onto the surfactant lamellae. The
temperature behavior provides a convenient and well-reproducible way
to initialize the dynamics, by rapidly quenching the sample from its
low-$T$ fluid phase to the solid phase
\cite{RamosPRL2001,RamosPRL2005,MazoyerPRL2006}. At room
temperature, the onion phase behaves mechanically as a gel, with a
storage modulus, $G' \sim 300$ Pa, that is nearly
frequency-independent in the range $5 \times 10^{-3}-10$ Hz and
roughly one order of magnitude larger than the loss modulus, $G''$.

\begin{figure}
\includegraphics{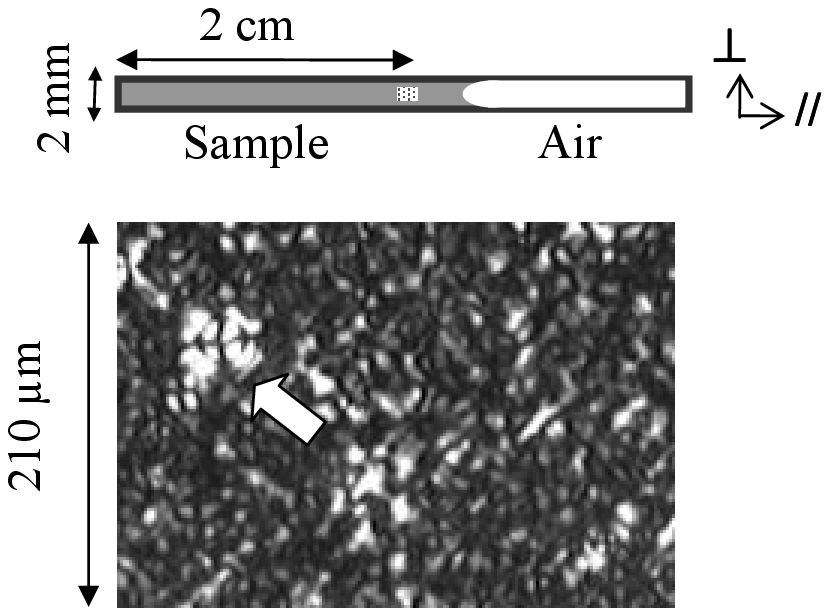}
\caption{(Top) Scheme of the capillary, with the air bubble, the
location of the field of view (dotted area), and the orientation of
the parallel and perpendicular axis. The size of the field of view
is $0.93 \, \rm{mm} \times 1.24 \, \rm{mm}$. (Bottom) Portion of a
typical image of the sample as taken by light microscopy between
cross-polarizers. The arrow points to a large onion: note the
characteristic four-lobed pattern due to the onion birefringence.}
\label{FIG:1}
\end{figure}

We probe the dynamics of the onion phase by means of optical
microscopy. The sample is contained in a glass capillary of
rectangular cross-section with thickness $200 \, \mu \rm{m}$ and
width $2 \, \rm{mm}$, which is flame-sealed in order to prevent
evaporation. Centrifugation is used to confine the air bubble at one
end of the capillary (see Fig.~\ref{FIG:1}). The sample is placed in
an oven (Instec) that controls the temperature $T$ to within
$0.15^{\circ} \rm{C}$ (standard deviation of $T$ over 1 day). $T$ is
measured with a thermocouple placed in direct contact with the
capillary. A light microscope equipped with a 10x objective is used
to image a thin slice of the sample, in the mid plane between the
upper and lower wall of the capillary. The field of view is $0.93 \,
\rm{mm} \times 1.24 \, \rm{mm}$ and is located in the center of the
capillary, as shown in the scheme of Fig.~\ref{FIG:1}. The onions
are difficult to visualize using conventional optical microscopy,
due to their weak refractive index mismatch with respect to the
solvent, and because they are densely packed. While individual
onions in diluted samples are clearly visible using differential
interference contrast microscopy~\cite{equilibriumonions},
closely-packed onions are best visualized under crossed polarizers.
Due to their birefringence and their spherical shape, they appear as
dark crosses on a bright background, as can be seen in
Fig.~\ref{FIG:1}. We use a charge-coupled detector camera (CCD) to
take images of the sample, with a delay between images of $15$ sec;
each image results from an average over $10$ frames, in order to
reduce the electronic and read-out noise of the CCD.

The sample is loaded in the capillary in its fluid state and the
capillary is immediately placed under the microscope, in the oven
regulated at $5^{\circ} \rm{C}$. The dynamics is then initialized by
a $T$ jump from $5^{\circ} \rm{C}$ to $23.3^{\circ} \rm{C}$, which
is performed \textit{in situ}. The time needed to reach the final
temperature is less than one minute. We define age $t_{w}=0$ as the
time at which $T$ has reached its stationary value $23.3^{\circ}
\rm{C}$. A typical experiment lasts about 1 day: we have checked
that no measurable drift of the capillary due to mechanical
instabilities occurs over this time.

\subsection{Image Analysis}

Due to the tight packing and to polydispersity, a typical onion
image consists of a highly contrasted intensity pattern, where only
a few, very large onions may be individually resolved, as shown in
Fig.~\ref{FIG:1}. Thus, it is impossible, in general, to track the
motion of individual objects. Instead, we use a method inspired by
the particle imaging velocimetry (PIV) technique widely used in
fluid mechanics to obtain coarse-grained displacements fields of
fluids seeded with tracer particles~\cite{WillertExpFluids1991}.
Similar approaches specialized to the case where individual
particles are not discernable, as in our case, have been termed
digital imaging correlation (DIC)~\cite{WesterweelMeasSciTech1997}
or image correlation velocimetry (ICV)~\cite{TokumaruExpFluids1995}.
In these methods, a time series of images of the sample is taken,
usually using a CCD camera. Each image is divided into a grid of
regions of interest (ROIs), corresponding to square regions of side
$L$ in the sample. The internal dynamics within a given ROI is
supposed to be negligible compared to the overall drift of that ROI.
In other terms, the variation of the displacement field over a
distance $L$ is supposed to be much smaller than its absolute
magnitude. Under this assumption, the intensity pattern of a given
ROI in an image taken at time $t$ will appear to be shifted in the
corresponding ROI imaged at time $t+\tau$. The displacement can be
estimated by calculating by what amount a ROI of the second image
has to be back-shifted in order to maximize its resemblance with the
corresponding ROI of the first image. By repeating this procedure
for all ROIs, a coarse-grained displacement field is obtained.

Usually, for a given ROI the shift along the direction of columns
($\delta x$) and rows ($\delta y$), is determined by searching the
maximum of the spatial crosscorrelation of the intensity of the two
images. With this method, the displacement is determined with a
resolution of one pixel. Several schemes have been proposed to
improve this resolution, e.g. by calculating the center of mass of
the peak of crosscorrelation, or by fitting it to a 2-dimensional
analytical function such as a Gaussian. We find that these standard
approaches do not perform very well in our case, most likely due to
the anisotropic shape of the correlation peak resulting from the
peculiar intensity pattern in the onion images. We have therefore
developed an alternative approach based on a least-square method,
where we minimize $\chi^2(\delta x,\delta y)$, the variance of the
difference between the first image, $I$, and a shifted version of
the second image, $J$. In order to calculate the optimum shift with
subpixel resolution, we use linear interpolation and define $\chi^2$
as follows:
\begin{eqnarray}
\chi^2(\delta x,\delta y)& = \sum_{r,c} \Delta_{r,c}^2(\delta
x,\delta y)
\\
\Delta_{r,c}(\delta x,\delta y)& = I_{r,c}-(1-\delta x)(1-\delta
y)J_{r,c}-
\\ \nonumber
& \delta x(1-\delta y)J_{r,c+1}-(1-\delta x)\delta yJ_{r+1,c}
\\ \nonumber & -\delta x\delta yJ_{r+1,c+1}
\, .
\end{eqnarray}
Here, $I_{r,c}$ is the intensity at time $t$ of the pixel at row $r$
and column $c$ and $J_{r,c}$ the corresponding quantity at time
$t+\tau$. The sum is over all pixels belonging to a given ROI. For
simplicity, Eq.~(2) has been written for the case $0 \le \delta x
\le 1 $, $0 \le \delta y \le 1$. The general case and the details of
the algorithm will be discussed in a forthcoming paper.

We have performed several tests of our algorithm. First, we have
checked that the gradient of the displacement field is not too high,
as required by PIV-based methods. We find that the difference of the
displacement field measured for adjacent ROIs is typically at most
20\% of the displacement itself, thus validating our approach.
Furthermore, we find that different choices of the ROI size yield
consistent results, as long as $L$ is at least a few onion
diameters. As one would expect, increasing the ROI size reduces the
measurement noise, but results in a coarser displacement field. The
best trade-off between high spatial resolution and low noise is
found for $L = 78~\mu\mathrm{m}$, corresponding to 47 pixels. The
full displacement field is then composed of $N = 192$ ROIs, on a 12
rows $\times$ 16 columns grid. To quantify the accuracy and the
precision of our measurements, we have used a piezoelectric motion
controller to repeatedly translate a sample by a known amount
($1~\mu\mathrm{m}$). The measured displacement field was found to
agree with the imposed one to within 80 nm in the worst case. In the
following, we will take this value as the typical uncertainty of our
measurements.

\section{Results}
\label{Results}
\subsection{Time-resolved dynamics}

The raw data obtained from the image analysis are two-time
displacement maps, $\Delta R_{\alpha}^{i,j}(t_{\mathrm w},\tau)$.
Here, $\alpha$ stands either for $\parallel$ or $\perp$ and refers
to the component of the displacement along or perpendicular to the
capillary long axis (see Fig.~\ref{FIG:1}). The superscript $i,j$
designate the $i$-th row and the $j$-th column of the ROI grid
($1\leq i \leq 12$ and $1\leq j \leq 16$), and the displacements are
measured between images taken at time $t_{\mathrm w}$ and
$t_{\mathrm w}+\tau$. Although in the following we will discuss the
behavior of both components, the focus will be on $\Delta
R_{\parallel}^{i,j}(t_{\mathrm w},\tau)$, since the dynamics along
the long axis of the capillary typically dominates the one in the
perpendicular direction.

In addition to the local dynamics, we also analyze time-resolved
global quantities, obtained by averaging over space. For each
component, we define $\Delta R_{\alpha}(t_{w},\tau)$, the spatially
averaged absolute displacement between time $t_{\mathrm w}$ and
$t_{\mathrm w}+\tau$:
\begin{equation}
 \Delta R_{\alpha}(t_{w},\tau) = \frac {1}{N} \sum_{i,j}
\Delta R_{\alpha}^{i,j}(t_{w},\tau)\, ,
\end{equation}
where the sum is over the $N$ ROIs of the grid. We also introduce
the corresponding spatially averaged relative displacement:
\begin{align}
 \Delta r_{\alpha}(t_{w},\tau)  = \sqrt{\frac {1}{N} \sum_{i,j} \left [\Delta
r_{\alpha}^{i,j}(t_{w},\tau)\right ]^2 }\, ,
\end{align}
where $\Delta r_{\alpha}^{i,j}(t_{w},\tau) = \Delta
R_{\alpha}^{i,j}(t_{w},\tau)- \Delta R_{\alpha}(t_{w},\tau)$ is
the local deviation with respect to the spatially averaged
dynamics. The relative displacements are by definition positive
and quantify the spatially heterogeneous character of the
instantaneous displacement field. The quantities defined above are
function of both the age of the sample, $t_{w}$, and the lag
between images, $\tau$. In Ref.~\cite{MazoyerPRL2006} $\Delta
r_{\parallel}(t_{w},\tau)$ has been studied as a function of
$t_{w}$: here, we fix the age of the sample and explore the
evolution of the absolute and relative displacements as a function
of the lag $\tau$ between images. Although most results presented
in the following refer to a specific age, we have checked that
they are representative of the general behavior of the sample.

\begin{figure}
\includegraphics {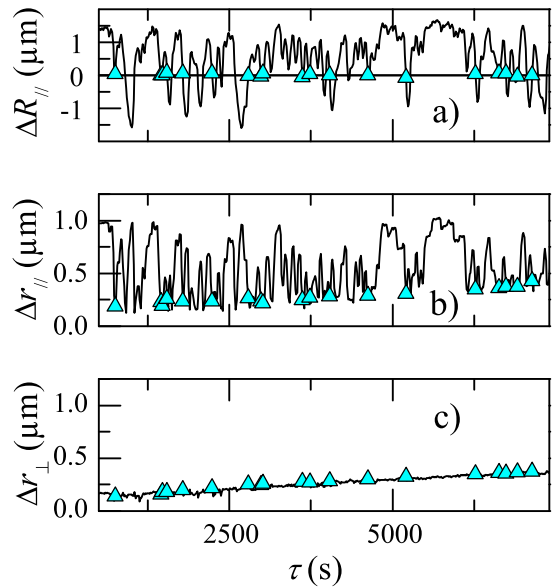}
\caption{(Color on line) The three panels show, for a sample age
$t_{w}= 15 \; 390$ s and as a function of lag $\tau$, the
spatially-averaged absolute displacement parallel to the capillary
axis (a) and the spatially-averaged relative displacement parallel
and perpendicular to the capillary axis (b and c, respectively).
The black lines are the data for all lags $500 \geq \tau \le 7400$
s, while the cyan triangles indicate irreversible events alone, as
defined in the text.} \label{FIG:2}
\end{figure}

We plot in Fig.~\ref{FIG:2}a the lag dependence of the parallel
absolute displacement, $ \Delta R_{\parallel}(t_{w},\tau)$, for an
age $t_{w}=15390$ s and a restricted, representative range of
lags. $ \Delta R_{\parallel}$ is temporally heterogenous, with
intermittent peaks, whose magnitude is of the order of $1$ $\mu$m.
The time between these events is of the order of a few hundreds
seconds. Neither the amplitude of the peaks nor their frequency
appear to evolve systematically with $\tau$. Indeed, the
displacement averaged over a time window significantly longer than
the typical time between peaks, but significantly shorter than the
duration of the experiment, does not vary with time (data not
shown). This indicates that the sample does not drift, and that
the dynamics is stationary. In Ref.~\cite{MazoyerPRL2006} we have
shown that the intermittent behavior is due to small,
experimentally unavoidable fluctuations of the temperature imposed
by the oven. These fluctuations are of the order of $0.1^{\circ}
\rm{C}$ and lead to elongations and contractions of the sample,
due to the thermal expansions of water, the main constituent of
our system. The fluctuations of the spatially averaged absolute
displacement $\Delta R_{\parallel}$ are quite peculiar. They
exhibit abrupt negative-going swings, corresponding to $\Delta T <
0$, followed by smoother positive displacements. These features
reflect the time evolution of $T$ due to the oven thermostat,
which cools abruptly while heating up more smoothly. Because of
the capillary geometry, $T$ fluctuations lead to an uniaxial
elongation or contraction of the sample, along the $\parallel$
axis. In agreement with a uniaxial thermally-induced
expansion/compression mechanism, we measure that the absolute,
spatially averaged,  displacement along the $\perp$ axis is always
very small and comparable to the measurement uncertainty (data not
shown).

We show in Fig.~\ref{FIG:2}b and~\ref{FIG:2}c the parallel and
perpendicular spatially averaged relative displacements, $\Delta
r_{\parallel}$ and $\Delta r_{\perp}$, as a function of the lag,
for the same lag window as that used in Fig.~\ref{FIG:2}a. The
small range of lags shown in Fig.~\ref{FIG:2} allow a close
comparison of the three signals. The perpendicular relative
displacement (Fig.~\ref{FIG:2}c) increases smoothly with time,
without any significant temporal fluctuation. By contrast, the
parallel component of the relative displacement
(Fig.~\ref{FIG:2}b) exhibits intermittent peaks, whose amplitude
is several tenths of $\mu$m. This intermittent behavior is similar
to that of the absolute parallel displacement shown in a).
In~\cite{MazoyerPRL2006} we have analyzed the same quantities for
a fixed lag, as a function of $t_{\mathrm w}$, and have shown that
the intermittent peaks of the relative and absolute displacement
are simultaneous and that both are induced by temperature
fluctuations. The same conclusions are reached here; note that
both positive and negative peaks of $\Delta R$ are associated to
positive spikes of $\Delta r$, since the latter is by definition
positive. The analysis of $\Delta R$ and $\Delta r$ \textit{vs}
$\tau$ presented here further clarifies the nature of the dynamics
of the onions: while the absolute displacement fluctuates around a
constant value \cite{NoteDeltaR}, the parallel and perpendicular
relative displacements overall increase with lag $\tau$. In other
words, even when the sample comes back to a previous position
\textit{on average}, as indicated by $\Delta R = 0$, the local
configuration has changed, since $\Delta r \neq 0$: this proves
directly the existence of irreversible rearrangement events. It is
worth noting that the scenario emerging from these observation is
reminiscent of the dynamics of dense emulsions or colloidal
suspensions to which a periodic shear is
applied~\cite{HebraudPRL1997,PetekidisPRE2002}, although the
physical origin of the motion is different. In those systems, the
particle positions decorrelate as a result of an applied shear
stress. In our sample, elongations and contractions due to
temperature fluctuations are not spatially uniform, as we will
discuss in more detail in Sec.~\ref{SEC:IIIb}. Therefore, they
result in a shear deformation analogous to that experienced by the
systems in~\cite{HebraudPRL1997,PetekidisPRE2002}. In the
experiments described in~\cite{HebraudPRL1997,PetekidisPRE2002},
not all emulsion droplets or colloidal particles recover their
initial position when the applied stress is removed, because some
plastic rearrangements have occurred. Similarly, the configuration
of our system is permanently changed after several
contraction/elongation cycles, because irreversible rearrangements
have occurred.

In order to characterize the irreversible rearrangement events,
for a given $t\mathrm{_{w}}$ we identify the lags for which the
absolute parallel displacement is null. For these lags, we expect
no contribution to the displacement field due to the shear induced
by contraction or elongation, since the sample is globally
undeformed. Hence, any local displacement can be attributed
exclusively to irreversible rearrangements. In practice, we
identify the lags $\tau$ for which $|\Delta R_{\parallel}| \le (0
\pm 0.08)~\mu$m, where the range of accepted absolute
displacements is chosen equal to the measurement uncertainty.
These lags are marked by triangles in Fig.~\ref{FIG:2}. We find
that all the symbols lie on the ``baseline'' of $\Delta
r_{\parallel}$, indicating that this procedure indeed identifies
only those lags for which the change in sample configuration is
minimal, i.e. lags for which no extra contribution due to a
transient shear is present.

\begin{figure}
\includegraphics {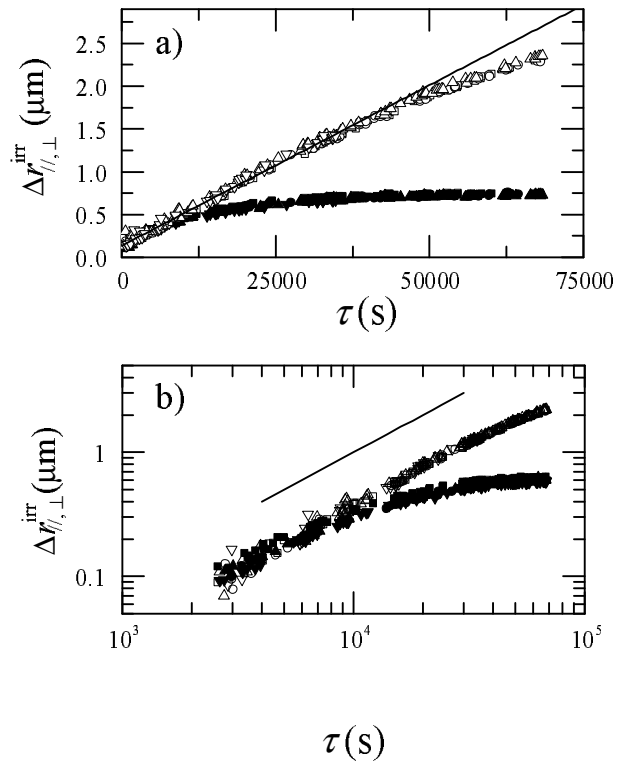}
\caption{(Color on line) $\tau$ dependence of the spatially-averaged
relative displacements, parallel (empty symbols) and perpendicular
(full symbols) to the capillary long axis, for the irreversible
events alone, as defined in the text. The four different symbols
correspond to data taken at close but different sample ages:
$t_{\mathrm w}=15030$ s (squares), $15390$ s (circles) , $15780$ s
(up triangles) and $16140$ s (down triangles). Lin-lin plot in a)
and log-log plot in b). In b) an offset on the order of $0.13 \, \mu
m$ has been subtracted to the data. The black line in a) is a linear
fit of the data for $\tau \leq 35000$ s. In b), the black line is a
power law with an exponent of $1$.} \label{FIG:3}
\end{figure}

\begin{figure}
\includegraphics {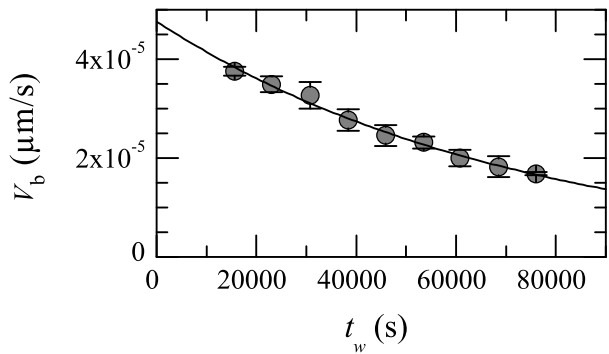}
\caption{(Symbols) Ballistic velocity for irreversible motion
along the parallel direction as a function of sample age. The
error bars are the standard deviations for analysis performed for
at least $4$ close but distinct sample ages. The line is an
exponential fit yielding a decay time of $(72000 \pm 3000)$ s.}
\label{FIG:4}
\end{figure}

We show in Fig.~\ref{FIG:3} the lag-dependence of the relative
displacements both parallel, $\Delta
r_{\parallel}^{\mathrm{\mathrm{irr}}}$, and perpendicular, $\Delta
r_{\perp}^{\mathrm{irr}}$, to the long axis of the capillary for
the irreversible events alone. Different symbols correspond to
data analysis performed for four close but distinct sample ages in
the range 15030 s~$\le t_{\mathrm w} \le$~16140 s. The four sets
of data nicely collapse onto two unique curves. Remarkably, we
find that, up to $\tau \approx 35000$ s, i.e. over more than a
decade in time, $\Delta r_{\parallel}^{\mathrm{irr}}$ varies
linearly with the lag.  A fit with a straight line, $\Delta
r_{\parallel}^{\mathrm{irr}} = V_{\mathrm{b}}(t_{\mathrm w})\tau +
b$, is shown in Fig.~\ref{FIG:3}a. We find $b = (0.13\pm 0.1)
\mu$m, of the order of the measurement uncertainty. Thus, up to
experimental uncertainties, the initial growth of $\Delta
r_{\parallel}$ is proportional to $\tau$: the ultraslow motion
associated with irreversible rearrangements is ballistic. For the
data of Fig.~\ref{FIG:3}, taken at sample ages from $t_{\mathrm
w}=15030$ s to $t_{\mathrm w}=16140$ s, we find $V_{\mathrm{b}}=
(3.8 \pm 0.1) \times 10^{-5} \mu\mathrm{m\,s}^{-1}$. The ballistic
nature of the dynamics for not too long lags is robust: indeed,
ballistic motion is measured for all sample ages for lags shorter
than $\tau \approx 35000$ s. As shown in Fig.~\ref{FIG:4},
$V_{\mathrm{b}}$ decreases decreases steadily with sample age,
$t_{\mathrm w}$. An exponential decay, with a characteristic decay
time of $(72 000 \pm 3000)$ s, can fit the data, although
experimental measurements over a larger range of sample ages would
be required to better characterize the functional form of the
decrease. In Ref.~\cite{MazoyerPRL2006} the contribution of the
irreversible events to the overall dynamics has been estimated
using a different method, which led to similar conclusions: the
irreversible dynamics is ballistic, with a characteristic velocity
that decays nearly exponentially with sample age. The decay time
of $V_{\mathrm{b}}$ found here is about twice as large as the one
($40 000 \pm 5500$ s) evaluated in~\cite{MazoyerPRL2006}, probably
due to the difference in the analysis method. The analysis of
$\Delta R$ and $\Delta r$ \textit{\textit{vs}} $\tau$ presented
above shows that the typical frequency of the
contracions/elongations causing the irreversible rearrangements
does not change with $t_{\mathrm w}$, as expected for a mechanism
driven by externally imposed temperature fluctuations. We thus
conclude that the aging revealed by the decrease of
$V_{\mathrm{b}}$ must be due to a decrease of the amplitude of the
displacement associated with each event.

The departure of $\Delta r_{\parallel}^{\mathrm{irr}}$ from a
straight line for lags larger than $\approx 35000$ s (see
Fig.~\ref{FIG:3}) is due to the slowing down of the dynamics, and
the decrease of the velocity with sample age. The slowing down of
the dynamics is more pronounced in the perpendicular direction than
in the parallel direction. Indeed, a saturation of $\Delta
r_{\perp}^{\mathrm{irr}}$ at a value of about $0.6 \, \mu m$ is
observed for lags larger than $\sim 35 000$ s, while no saturation
is observed for $\Delta r_{\parallel}^{\mathrm{irr}}$ over the
duration of the experiment.

\subsection{Individual trajectories and spatial structure of the dynamics}
\label{SEC:IIIb}

\begin{figure}
\includegraphics {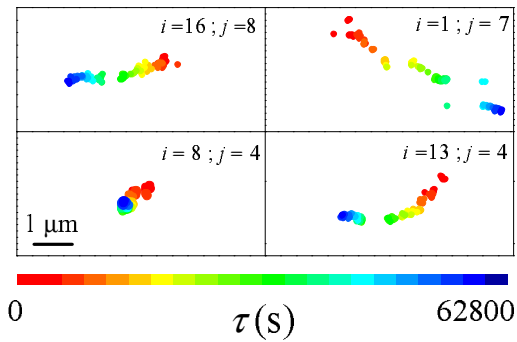}
\caption{(Color online) Trajectory of $4$ single ROIs with respect
to the position at time $t_{\mathrm{w}}=15390$ s. The location of
each ROI on the 12 $\times$ 16 overall grid is specified by the row
and column indices $i$ and $j$. Only data points corresponding to
irreversible events as defined in the text are plotted. The color of
the circles changes according to the delay $\tau$, as shown by the
scale. The spatial scale is the same for the four trajectories.}
\label{FIG:5}
\end{figure}

The ballistic motion discussed so far has been inferred from the
behavior of $\Delta r(t_{\mathrm w},\tau)$, a quantity that is
spatially averaged. To better characterize the irreversible events,
we analyze also the time evolution of the local relative
displacements. To this end, we study the trajectory of individual
ROIs, where only delays corresponding to irreversible events are
considered. Representative trajectories located at different
positions in the grid of ROIs are shown in Fig.~\ref{FIG:5}, where
the displacements relative to the position at $t_{\mathrm w} =
15390$ s are plotted as color-coded circles, according to the lag
$\tau$. We find that all trajectories are close to straight lines,
the main changes between different ROIs being the length of the
trajectory and its orientation, i.e. the magnitude and the
orientation of the local velocity.

\begin{figure}
\includegraphics {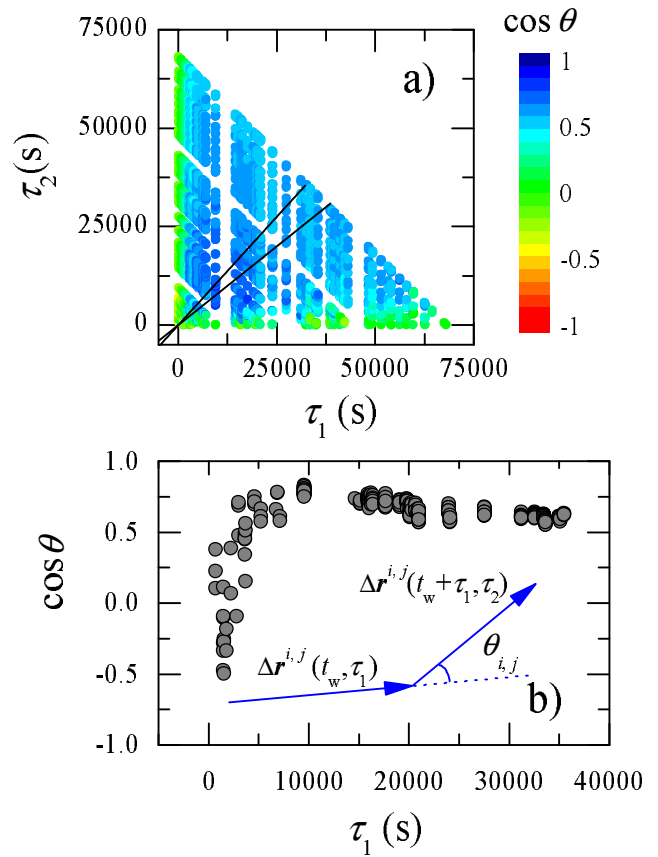}
\caption{(Color online) a) Color map representation of
$\cos\theta(\tau_1,\tau_2)$, with $\theta$ the angle between
displacements over consecutive time intervals of duration $\tau_1$
and $\tau_2$ (see text for details). b) Same data for
$\tau_2=\tau_1\pm 20 \%$, as a function of $\tau_1$. Inset:
schematic definition of $\theta_{i,j}$.} \label{FIG:6}
\end{figure}

To quantify our observations, we evaluate for all ROIs the
correlation between the direction of the relative displacement
between times $t_{\mathrm w}$ and $t_{\mathrm w}+\tau_1$ and that
between times $t_{\mathrm w}+\tau_1$ and $t_{\mathrm
w}+\tau_1+\tau_2$ (see inset of Fig.~\ref{FIG:6}b). More
specifically, we calculate
\begin{eqnarray}
\cos\theta^{i,j}(\tau_1,\tau_2)=\frac{\Delta
\textbf{r}^{i,j}(t_{\mathrm w},\tau_1) \cdot \Delta
\textbf{r}^{i,j}(t_{\mathrm w}+\tau_1,\tau_2)}{|\Delta
\textbf{r}^{i,j}(t_{\mathrm w},\tau_1)|
 | \Delta \textbf{r}^{i,j}(t_{\mathrm w}+\tau_1,\tau_2)|}
\end{eqnarray}
and its average over all ROIs, $\cos\theta$. If the trajectories
were random walks, $\cos\theta^{i,j}$ would be distributed between
$-1$ and $1$, and its average value would be close to zero. By
contrast, for ballistic motion we expect $\cos \theta$ close to
$1$. We plot in Fig.~\ref{FIG:6}a the value of $\cos \theta$, in
the plane $(\tau_2, \tau_1)$. The color-coded map shows that,
except for small $\tau_1$ or small $\tau_2$ for which results are
dominated by experimental uncertainties, the value of $\cos
\theta$ is always larger than $0.5$, suggesting a strong
correlation of the direction of motion over long time scales.

We now consider the case  of two successive displacements with
approximately the same duration, hence $\tau_2\approx \tau_1$. We
thus show in Fig.~\ref{FIG:6}b we show $\cos\theta$ for
$\tau_2\approx \tau_1$. The data are obtained from the points in
Fig.~\ref{FIG:6}a that lie in between the two straight lines
$\tau_2=\tau_1\pm 20 \%$ shown in the figure. At short lags, for
which $\Delta r_{\parallel}^{\mathrm{irr}}$  and $\Delta
r_{\perp}^{\mathrm{irr}}$ are not significantly different from the
experimental uncertainties, $\cos \theta$ varies randomly between
$-0.5$ and $0.5$ before reaching rapidly a plateau of height $0.67
\pm 0.06$ for $\tau_1 > 4500$ s, corresponding to an average angle
of $48$ deg. The maximum value for $\cos \theta$ is $\approx 0.8$,
for $\tau_1 \simeq 10 000$ s (corresponding to an angle $\theta$
of $37$ deg). The slightly decreasing trend above $10 000$ s is
due to the saturation of $\Delta r_{\perp}^{\mathrm{irr}}$ at
large $\tau$ (Fig.~\ref{FIG:3}), which is associated with a kink
in individual trajectories, as shown in Fig.~\ref{FIG:5}. Overall,
$\cos\theta$ is close to unity, indicating that the direction of
motion persists over long times (as long as the duration of the
experiment). This analysis demonstrates that the motion is
ballistic not only when averaged over the whole field of view, but
also at a local level, since most individual ROIs move in a well
defined direction and exhibit trajectories close to straight line
over the whole duration of the experiment.

\begin{figure}
\includegraphics {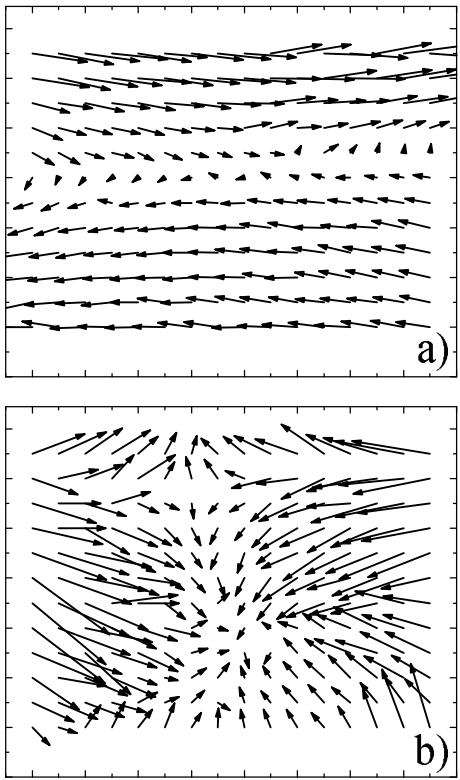}
\caption{a) Representative example of a shear-like relative
displacement field, observed for $\tau=615$ s and $t_{w}=15390$ s.
b) Vortex-like relative displacement field, observed for
$\tau=68100$ s and $t_{w}=15390$ s. The field of view is
$0.93~\mathrm{mm} \times 1.24~\mathrm{mm}$; the arrows are
magnified $50$ times.} \label{FIG:7}
\end{figure}

What is the spatial structure of the displacement field associated
with such ballistic dynamics? By examining a large number of
displacement fields for all ages and all delays, we find that
generally they can be divided in two classes. The first one
corresponds roughly to a shear deformation along the long axis of
the capillary, as shown in Fig.~\ref{FIG:7}a. This displacement
pattern is associated with a significant elongation or contraction
of the sample; thus it corresponds to the transient, reversible
peaks of the relative displacement. The second class is
characterized by a vortex-like (or compression-like) structure, as
seen in Fig.~\ref{FIG:7}b. Note that, since the sample is actually
incompressible, Fig.~\ref{FIG:7}b suggests that some motion may
occur in the direction perpendicular to the plane that is imaged.
In contrast to the shear-like deformation, this pattern is quite
isotropic and implies non-negligible displacements along the
$\perp$ direction. The events that have been previously identified
as irreversible are typically associated with a relative
displacement field of the second kind. The overall pattern of the
displacement fields associated with irreversible events does not
significantly vary with time, as expected from the analysis of the
individual trajectories, which are all very close to straight
lines.

From our analysis, we can conclude that the $T$-induced,
intermittent events and the ballistic irreversible rearrangements
can be associated to distinct, specific structures of the relative
displacement field. The former correspond to the (anisotropic)
shear deformation of the sample  along the capillary axis
(Fig.~\ref{FIG:7}a), while the latter correspond to nearly
isotropic, compression-like displacement fields
(Fig.~\ref{FIG:7}b). Both structures are correlated over distances
comparable to the field of view. We have checked that they do not
exhibit significant changes with sample age, as expected from the
ballistic motions of the individual ROIs.

\section{Discussion and conclusion}

We have used optical microscopy to study the slow dynamics of
closely packed onions, a soft glassy material. Two classes of
dynamical events have been identified: reversible and irreversible
rearrangement. We have shown that the reversible, intermittent peaks
of relative displacement are due to the contraction or elongation of
the sample, due to temperature fluctuations, and correspond to a
shear deformation along the long axis of the capillary. The origin
of the shear associated with thermal expansion or contraction
remains somehow unclear. It may be due to the curvature of the
meniscus at the interface between the free end of the sample and the
air bubble. It may also stem from small thermal gradients in the
sample, close to the capillary wall. Indeed, portions of the sample
close to the lateral wall are in better thermal contact with the
oven, with respect to the central region. Therefore, they may react
faster to any small change in $T$ and lead to a larger elongation or
contraction, as we observe in our experiments. We thus expect the
precise spatial structure of the displacement fields to depend on
the sample geometry (e.g. to be different in a microscopy capillary,
 a light scattering cuvette, and a Couette cell for rheology measurements),
 but the effect of $T$ fluctuations
should be relevant in all experimental configurations.

The second class of dynamical events are the irreversible
rearrangements, presumably occurring as the result of repeated shear
cycles. They correspond to a nearly isotropic displacement field
that is spatially correlated over very large distances. The motion
resulting from these ultraslow rearrangements is ballistic, with a
velocity that decreases exponentially as the sample ages. We believe
that these irreversible events occur at all times, presumably even
when they are masked by larger, transient shear events caused by
temperature fluctuations.

Below, we discuss shortly the main features of the irreversible
rearrangements and finally comment on the role of the temperature
fluctuations as the driving force for the slow dynamics.

\subsection{Ballistic motion and aging}

We have shown that the amplitude of the relative displacements is
proportional to the lag, $\tau$, hence proving directly the
ballistic nature of the ultraslow motion. A similar ballistic
motion has been invoked to explain the slow dynamics of a variety
of soft glassy materials probed by light and X-photon scattering
techniques
\cite{LucaFaraday2003,RamosPRL2001,BellourPRE2003,BandyopadhyayPRL2004,ChungPRL2006,RobertEPL2006,GuoPRE2007},
including the very same onion sample investigated
here~\cite{RamosPRL2001}. To the best of our knowledge, our
experiments are the first proof in direct space of ballistic
motion in the slow relaxation of a glassy material. Below, we
provide simple scaling arguments to understand the order of
magnitude of the ballistic velocity measured experimentally.

In Ref.~\cite{RamosPRL2005}, we have presented a model to
rationalize the dynamical behavior of the onion glass, speculating
that motion results from the combined action of viscous forces and
internal elastic stress. We briefly recall the main features of the
model: regions of size $D$ containing several onions move under the
action of elastic forces due to the relaxation of internal stresses
and experience a viscous drag. The driving force due to the
relaxation of internal stress is $F_{\sigma}=\sigma D^{2}$, where
$\sigma =\varepsilon G$ is the elastic stress, $G$ the elastic
modulus, and $\varepsilon$ the strain. Stokes' law gives the viscous
force: $F_{\eta}=p \eta D v$ where $p$ is a geometrical factor of
order $10$ that depends on the shape of the region, $\eta$ the
effective viscosity of the medium, and $v$ the velocity of the
region. By balancing the viscous and the elastic force and solving
for $v$, one obtains
\begin{equation}
 v =G  \varepsilon D/(\eta p) \, . \label {EQ:stress}
\end{equation}
We calculate $v$ using experimental values for $\eta$,
$\varepsilon$, $G$ and $D$. To estimate the order of magnitude of
the strain, we consider the irreversible displacement fields
measured experimentally and take the difference of the relative
displacement between a ROI for which it is close to $0$ and one for
which it is maximum (about $1 \, \mu$m for $t_{\mathrm w}=15000$ s).
The typical distance between these two ROIs is about
$500~\mu\mathrm{m}$, yielding $\varepsilon \approx 0.002$. In our
microscopy measurements, we probe the motion of ROIs of size $L$, so
we set $D = L = 78 \, \mu\mathrm{m}$. We take for $\eta$ the value
of the effective viscosity deduced from step strain relaxation
measurements~\cite{RamosPRL2005}, $\eta=3.4 \times
10^{5}~\mathrm{Pa\, s}$ for an age $t_{\mathrm w}=15000$ s, and take
$G \simeq 300$ Pa as measured by linear
rheology~\cite{equilibriumonions}. Using Eq.~(\ref{EQ:stress}), we
then find $v \approx 10^{-10}/p~\mathrm{m\,s}^{-1}$ with $p$
typically of the order of ten. This result is in very good agreement
with $V_\mathrm{b}(t_{\mathrm w}=15000~\mathrm{s}) = 3.8 \times
10^{-11}~\mathrm{m\,s}^{-1}$ measured by light microscopy.

It is interesting to compare $V_\mathrm{b}$ to $V_{\mathrm{DLS}}$,
the ballistic velocity obtained in~\cite{RamosPRL2005} from the
analysis of DLS data for the same sample. For an age $t_{w}=15000$ s
we found in~\cite{RamosPRL2001} $V_\mathrm{DLS} = 5 \times
10^{-13}~\mathrm{m\,s}^{-1}$, about two orders of magnitude lower
than $V_\mathrm{b}$ measured, at the same age, by light microscopy.
Several factors
may be invoked to explain the large difference. First, the dynamics is not probed on
the same length scale in the two experiments: here, the relevant
length scale is $L =78 \, \mu\mathrm{m}$, the ROI size. In the DLS
experiment, by contrast, the motion is probed on a much smaller
length scale, set by the inverse scattering vector and of the order
of $0.2-1~\mu\mathrm{m}$. Furthermore, the microscopy experiments
point to the role of $T$-induced strain fields as the driving force
for the slow rearrangements. In this scenario, the latter most
likely vary with both the frequency and the amplitude of the
transient strain, which depend on experimental conditions such as
the cell size and geometry and the temperature control, different in
microscopy and DLS measurements. More experiments, e.g.
varying the cell geometry, would be useful to clarify this issue.

The slowing down of the dynamics with sample age, a feature typical
of glassy dynamics, is reflected here by the decrease of
$V_\mathrm{b}$ with $t_{\mathrm w}$. $V_\mathrm{b}$ is a global
quantity, averaged over both space and time. Its decay may a priori
stem from one or more distinct mechanisms: a decrease of the rate at
which rearrangements occur, a reduced average displacement
associated with each rearrangement, or a smaller size of the volume
affected by a single rearrangement. Our space- and time resolved
measurements rule out the former and the latter hypothesis:
rearrangements occur at the same rate, imposed by the characteristic
time of the temperature fluctuations of the oven, and affect
essentially the same (large) portion of the sample at all ages.
Experiments on concentrated colloidal hard
spheres~\cite{CourtlandJPCM2003} have reached a somehow similar
conclusion: the size of the ``clusters'' formed by the fastest
moving particles did not evolve significantly during aging (Note
that these clusters contained just a few particles, in contrast to
the very long-ranged dynamical correlations observed here). A few
simulation studies have addressed the issue of the age dependence of
the correlation length of the dynamics, yielding a different answer:
the correlation length was found to grow with $t_{\mathrm w}$ both
in a Lennard-Jones~\cite{CastilloNatPhys2007,Parseiancondmat2006}
and a spin glass~\cite{KiskerPRB1996,BerthierPRB2002}. More
experiments and simulations will be needed to understand these
discrepancies.

\subsection{Long-ranged spatial correlations of the slow dynamics}

We have shown that the relative displacement fields associated with
the irreversible events exhibit surprisingly long-ranged spatial
correlations, comparable to the size of the field of view, thus much
larger than the onions' size. This result is strikingly different
from previous experimental and numerical determinations of the size
of dynamical heterogeneity, which generally are of the order of a
few particle sizes, e.g. in dense colloidal hard
spheres~\cite{WeeksScience2000,WeeksJPCM2007,BerthierScience2005},
granular
materials~\cite{DauchotPRL2005,DurianNatPhys2007,LechenaultEPL2008},
and molecular glass
formers~\cite{ReviewEdiger,ReviewGlotzer,CastilloNatPhys2007,BerthierPRE2007}.
Recent simulations in a two-dimensional hard sphere system may
however reconcile these findings \cite{BritoEPL2006,BritoJSTAT2007}.
The authors show that the rearrangements are short-range correlated
for supercooled, equilibrated samples, while they become correlated
over long distances deep in the out-of-equilibrium glass phase,
approaching the maximum packing. While previous works where
short-ranged correlations were found generally dealt with
equilibrium supercooled samples, it is clear that our
closely-packed, aging onions are jammed systems, for which
long-ranged correlations may be expected according
to~\cite{BritoEPL2006,BritoJSTAT2007}. Indeed, we have recently
observed very long-ranged spatial correlations in the dynamics of a
variety of jammed systems using space-resolved dynamic light
scattering~\cite{ThesisAgnes}, consistently with the scenario
emerging here.

As a further intriguing analogy, we note that the displacement field
following a rearrangement event in~\cite{BritoJSTAT2007} has a
vortex-like structure that is reminiscent of that observed in our
experiments. Similar structures have been also reported for the
non-affine part of the displacement measured, experimentally and
numerically, in various amorphous systems under shear, ranging from
granular material to glasses
\cite{Barrat1,Barrat2,KolbPRE2004,MaloneyPRL2006,DelGadoPRL2008}.
Whether these analogies are accidental or not remains an open
question.

\subsection{Role of the temperature fluctuations}

The experiments described in this paper highlight the key role of
temperature fluctuations in determining the slow dynamics of the
onions. One may wonder whether $T$-induced shear deformations are
the only driving force for the non-stationary irreversible
rearrangements. Indeed, it is difficult to imagine that thermal
energy alone could promote such a large scale motion. In fact, the
onions are a closely-packed, fully jammed system with a solid-like
behavior. In this respect, they are similar to dry foams or
compressed emulsions, none of witch exhibits a spontaneous slow
dynamics in the absence of a (non-thermal) driving force, such as an
applied stress or the stress continuously build up by
coarsening~\cite{DurianScience,AddadPRL2001}. For jammed or nearly
jammed systems, on the other hand, an applied shear strain or stress
has been shown to promote irreversible rearrangements in a variety
of systems, including granular
media~\cite{Pouliquen,Dauchot,ReisPRL2007},
emulsions~\cite{HebraudPRL1997}, and concentrated suspensions of
colloidal~\cite{ViasnoffPRL2002} and
non-colloidal~\cite{PineNatureShear,PineNaturePhysicsShear}
particles.

A final question concerns the generality of the mechanism reported
here. In our time-resolved light scattering experiments on dilute
colloidal gels~\cite{DuriEPL2006}, we did not observe any
correlation between temperature fluctuations and the rearrangement
events that were detected. By contrast, a strong correlation has
been established for the onion glass and has also been observed in
time-resolved light scattering measurements on a concentrated
micellar phase~\cite{ThesisAgnes}. We speculate that
temperature-induced strain fluctuations may be important in very
concentrated, solid-like systems where expansions and contractions
necessarily displace all constituents, and where thermal energy
alone may be insufficient to significantly rearrange the structure.
On the other hand, in dilute systems such as the gels, expansions
and contractions of the solvent would be accommodated flowing
through the porous structure, without affecting the particles. More
experiments with a finer temperature control or an externally
imposed shear will be necessary to test these ideas.

We thank M. Wyart for useful discussions. This work was supported in
part by the European NoE ``SoftComp'' (NMP3-CT-2004-502235) and the
French ACI JC2076. L.C. is a junior member of the Institut
Universitaire de France, whose support is gratefully acknowledged.


\begin{references}
\bibitem{NoteDeltaR} $\Delta R$ compares the sample configurations at time $t_{\mathrm w}$ and $t_{\mathrm w} + \tau$.
If the first image considered (at $t_w$) has been taken at high
temperature (with respect to the average temperature), then the
reference sample configuration is expanded, and $\Delta R$ will be
on average more compressed than the reference state and will thus
fluctuate with time around a negative value. By contrast if the
first image considered has been taken at relatively low
temperature, then the reference sample configuration is
``compressed'', and $\Delta R$ will be on average more expanded
than the reference state and will thus fluctuate with time around
a negative value. If the reference state corresponds to a relaxed
state, then $\Delta R$ will fluctuate around $0$.
\end{references}
\end{document}